\documentclass[journal=nalefd, manuscript=letter]{achemso}

\usepackage[version=3]{mhchem} 
\usepackage[T1]{fontenc}       
\usepackage{subfigure}
\usepackage{graphicx}
\usepackage{bm}
\usepackage{color}

\author{Wenhan Zhang}
\affiliation{Department of Physics and Astronomy, Rutgers University, Piscataway, New Jersey 08854, USA}
\author{M. X. Chen}
\affiliation{College of Physics and Information Science, Hunan Normal university, Changsha, Hunan 410081, China}
\author{Jixia Dai}
\affiliation{Department of Physics and Astronomy, Rutgers University, Piscataway, New Jersey 08854, USA}
\author{Xueyun Wang}
\affiliation{Department of Physics and Astronomy, Rutgers University, Piscataway, New Jersey 08854, USA}
\alsoaffiliation{School of Aerospace Engineering, Beijing Institute of Technology, Beijing 100081, China}
\author{Zhicheng Zhong}
\affiliation{Key Laboratory of Magnetic Materials and Devices, Ningbo Institute of Materials Technology and Engineering,  Chinese Academy of Sciences, Ningbo 315201,  China}
\author{Sang-Wook Cheong}
\affiliation{Department of Physics and Astronomy, Rutgers University, Piscataway, New Jersey 08854, USA}
\alsoaffiliation{Rutgers Center for Emergent Materials, Rutgers University, Piscataway, New Jersey 08854, USA}
\author{Weida Wu}
\email{wdwu@physics.rutgers.edu}
\phone{+1 848-445-8751}
\fax{+1 732-445-4343}
\affiliation{Department of Physics and Astronomy, Rutgers University, Piscataway, New Jersey 08854, USA}

\title{Topological phase transition with nanoscale inhomogeneity in (Bi$_{1-x}$In$_x$)$_2$Se$_3$}

\begin{document}
\newpage
\begin{abstract}
Topological insulators are a class of band insulators with non-trivial topology, a result of band inversion due to the strong spin-orbit coupling. The transition between topological and normal insulator can be realized by tuning the spin-orbit coupling strength, and has been observed experimentally. However, the impact of chemical disorders on the topological phase transition was not addressed in previous studies. Herein, we report a systematic scanning tunneling microscopy/spectroscopy and first-principles study of the topological phase transition in single crystals of In doped Bi$_2$Se$_3$. Surprisingly, no band gap closure was observed across the transition. Furthermore, our spectroscopic-imaging results reveal that In defects are extremely effective ``suppressors'' of the band inversion, which leads to microscopic phase separation of topological-insulator-like and normal-insulator-like nano regions across the ``transition''. The observed topological electronic inhomogeneity demonstrates the significant impact of chemical disorders in topological materials, shedding new light on the fundamental understanding of topological phase transition.
\end{abstract}

\newpage
Disorders are inevitably present in any functional materials or physical system.  Often, they have significant impact on many areas of physics, including condensed-matter physics~\cite{Ashcroft1976}, photonics~\cite{Wiersma2013,Segev2013}, and cold atoms~\cite{Shapiro2012}. E.g., quench disorders induce both frustration and randomness in doped magnets, the key ingredients of spin glass physics~\cite{Binder1986}. Quenched disorders due to chemical doping are associated with the observed nanoscale inhomogeneity in high $T_c$ cuprates~\cite{Pan2001, Pasupathy2008} and CMR manganites~\cite{Tomioka2004}. Strong disorders can also localize itinerant electrons~\cite{Lattices1956}, which is crucial for many appealing phenomena such as metal insulator transition and quantum Hall effects~\cite{Ando1975,Klitzing1980,Thouless1982}. However, the realization of quantum Hall effect requires high magnetic field, a hurdle for its broader applications. The quest for quantum Hall state without external magnetic field leads to development of quantum anomalous Hall~\cite{Haldane1988PRL}, quantum spin Hall~\cite{Bernevig2006Science, Konig2007Science}, which eventually lead to the birth of 3D topological insulator (TI)~\cite{Fu2007PRL, Kane2010RMP, Moore2010, Qi2011}.

In 3D TI, the non-trivial topology of band structure is a result of band inversion due to strong spin-orbital coupling (SOC)~\cite{Fu2007PRL, Kane2010RMP, Moore2010, Hasan2011, Qi2011}.  The change of topology class at the interface between topological and normal insulators ensures the existence of metallic Dirac surface states~\cite{Fu2007PRB, Zhang2009NatPhys, Xia2009, Hsieh2009a, Hsieh2009b, Chen2009}. Similarly, the topological distinction also enforces a zero band gap Dirac semi-metal state for spatially uniform topological phase transition (TPT). In theory, TPT can be induced by gradual change of average SOC strength via chemical doping, and has been experimentally established in TlBi(S$_{1-x}$Se$_{x}$)$_2$ and (Bi$_{1-x}$In$_x$)$_2$Se$_3$ via bulk measurements~\cite{Xu2011, Sato2011, Brahlek2012, Wu2013a, Lou2015,  Rao2015}. Because of topological stability, the band inversion is robust against weak disorders. The impact of strong disorders is quite rich. On one hand, strong lattice disorder can drive a TI into a normal insulator (NI)~\cite{Brahlek2016}. On the other hand, disorders could transform a metal with strong SOC into a topological Anderson insulator~\cite{Guo2010, Groth2009, Li2009}, though more elaborate theoretical analysis suggests that it is adiabatically connected to clean TI~\cite{Prodan2011}. Although TPT has been studied in few systems, the impact of disorders due to chemical doping has not been addressed. Furthermore,  Lou \textit{et al} reported a mysterious sudden band gap closure across TPT in In doped Bi$_2$Se$_3$ using angle resolved photoemission spectroscopy (ARPES)~\cite{Lou2015}. In this letter, we report systematic high-resolution scanning tunneling microscopy/spectroscopy (STM/STS) studies on single crystals (Bi$_{1-x}$In$_x$)$_2$Se$_3$, where $x$ varies from 0.2\% to 10.8\%. To our surprise, no band gap closing and reopening was observed across the TPT ($x_c\approx$ 6\%). Instead, the average band gap gradually increases across the transition. STS imaging reveals that In defects are extremely effective on changing topological properties locally, {\it i.e.} suppressing the topological surface states (TSS) and increasing band gap. The characteristic length of suppression is comparable with the decay length of TSS in Bi$_2$Se$_3$ ($\sim$1~nm), suggesting that individual In defects are effectively point NIs. The suppression is stronlgy enhanced in neighboring In defects, resulting in nanoscale mixture of TI and NI regions with proliferation of TSS inside the bulk across the TPT. The observation of TPT with nanoscale inhomogeneity will not only promote future investigations on the impact of disorders on TPT, but also open a door to optical sensing by harnessing percolative network of TSS.

High-quality single crystals of (Bi$_{1-x}$In$_x$)$_2$Se$_3$ with various In concentrations ($x$) were grown by self-flux method. An Omicron LT-STM with base pressure of 1$\times10^{-11}$\,mbar was used for STM/STS measurements. Electrochemically etched W tips were characterized on clean single crystal Au(111) surface before STM measurement~\cite{Chen1998}. Samples were cleaved in UHV at room temperature then immediately loaded into the cold STM head. All STM data were taken at 4.8~K. The differential conductance ($dI/dV$) measurements were performed with standard lock-in technique with modulation frequency $f=455$~Hz and amplitude $V_{mod}=5\sim 10$~mV.

The density functional theory (DFT) calculations were carried out using the Vienna \textit{ab initio} Simulation Package~\cite{Kresse1996,Kresse1996a}. A slab in 3$\times$3 supercell consists of 5\,QL terminated by Se was used to model the surface. The In doping ($\sim$1.1\%) is realized by substituting one of the Bi atoms in the sub-top layer with an In atom. The exchange correlation functional is approximated by the generalized gradient approximation as parametrized by Perdew, Burke and Ernzerhof~\cite{Perdew1996}, and pseudopotentials were constructed by the projector augmented wave method~\cite{Blochl1994,Kresse1999}. The 2D Brillouin zone is sampled by a 4$\times$4$\times$1 Monkhorst-Pack mesh. STM simulations were performed using the Tersoff-Hamann method~\cite{Tersoff1985}. Local density of states (LDOS) are calculated by the layer projection method~\cite{Chen2014a}, which integrates wave functions in spatial windows in vacuum.

\begin{figure*}[ht]
\centering
\includegraphics[width=\textwidth]{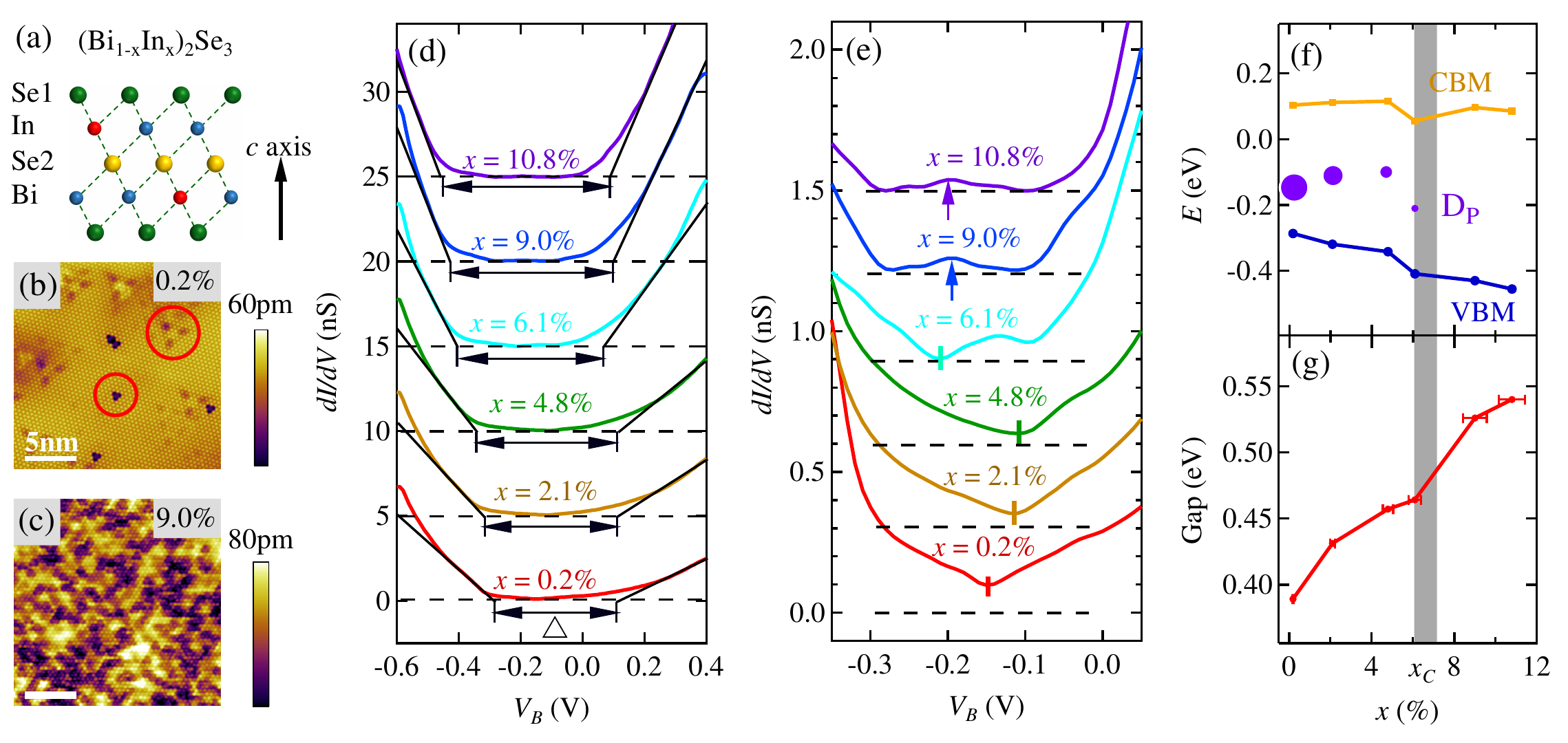}
\caption{(a) Schematic of crystal structure of 1~QL of (Bi$_{1-x}$In$_x$)$_2$Se$_3$. Se, Bi and In atoms are denoted by green/yellow, blue and red spheres, respectively. (b)-(c) STM topographic images of $x= 0.2$~\% and 9.0~\%. Tunneling condition: $-0.6$~V, 1~nA. (d) $dI/dV$ spectra of samples with different $x$ (offset for clarity). Black lines indicate the estimation of band edges. (e) Zoom-in $dI/dV$ spectra inside the band gap of (d). Dashed lines indicate the zero reference point and vertical bars mark the Dirac points. (f) VBM (blue), CBM (yellow) and Dirac point (violet) vs. $x$. The size of Dirac point symbols represent the spectra weight at Dirac point. (g) Estimated gap size at various $x$ showing a gradual increase of average band gaps. $x_c$: the critical point of TPT.}
\label{fig1}
\end{figure*}

Figure\,\ref{fig1}(a) shows the side view schematic of one quintuple layer (QL) of (Bi$_{1-x}$In$_x$)$_2$Se$_3$~\cite{Zhang2009NatPhys,Jiang2012PRL}. In atoms (red) preferentially occupy the Bi sites (blue). Atomically resolved topographic images were obtained in all the samples. The representative STM images of $x = 0.2$~\% and 9.0~\% are shown in Fig.\,\ref{fig1}(b) and \ref{fig1}(c) (See section 1 in Supporting Information for other samples). In atoms were observed in either the second or fourth atomic layer from surface. Fig.\,\ref{fig1}(d) shows the average $dI/dV$ spectra ($-$0.6 to 0.4\,eV) of each sample (offset vertically for clarity). All the spetra are obtained by averaging over 6000 single-point spectra in the sampling areas typically over 225~nm$^2$. The reference levels are marked by black dashed lines. Tunneling spectrum $dI/dV$ is proportional to the LDOS. Due to the presence of TSS, the LDOS within the band gap is not zero. Fig.\,\ref{fig1}(e) shows the zoom-in $dI/dV$ spectra inside the band gap. The Dirac point and the linear LDOS of TSS are clearly resolved in the $dI/dV$ spectra for $x\leq$ 6.1\%, but are less visible for $x>6.1\%$, indicating that the critical In concentration ($x_c$) of TPT is $\sim$ 6\%, in good agreement with prior transport and THz studies~\cite{Brahlek2012, Wu2013a}. Interestingly, the spectral weight of Dirac point (indicated by size of the symbol) gradually decreases as $x$ increases. Although it is not straightforward to determine the band gap $\Delta$ using STS in the presence of TSS,  the slope change of $dI/dV$ curves near band edges provides a reasonable criterion for estimation of conduction band minimum (CBM) and valence band maximum (VBM). The band edges are defined by the intercepts with the horizontal axis of the best linearly fitted $dI/dV$ near band edges, as shown with the black solid lines in Fig.\,\ref{fig1}(d) (see detailed description in Supporting Information). Similar results were obtained using a different criterion ({\it i.e.} ``kinks'' in the $d^2I/dV^2$ curves, see Supporting Information)~\cite{Xu2014}, corroborating the aforementioned method. Fig.\,\ref{fig1}(f) shows the the estimated CBM and VBM with the Dirac energy $E_D$ as functions of $x$. The CBM remains almost unchanged as $x$ increases while the VBM gradually shifts down. As shown in Fig.\,\ref{fig1}(g), this results in a monotonic increase of band gap toward that of  In$_2$Se$_3$ ($\Delta \sim1.3$~eV)~\cite{Brahlek2012}. Such behavior is in apparent conflict with the aforementioned picture of band closing and reopening, but is consistent with recent ARPES studies~\cite{Lou2015}.  Most of previous TPT studies assume a spatially uniform change of SOC with chemical doping. In reality, chemical doping inevitably introduces spatial inhomogeneity ({\it e.g.}, SOC), so TPT might not happen in a spatially uniform manner. To illustrate this, we performed systematic nanoscale spectroscopic mapping across the TPT.

\begin{figure}[ht]
\centering
\includegraphics[width=0.46\columnwidth]{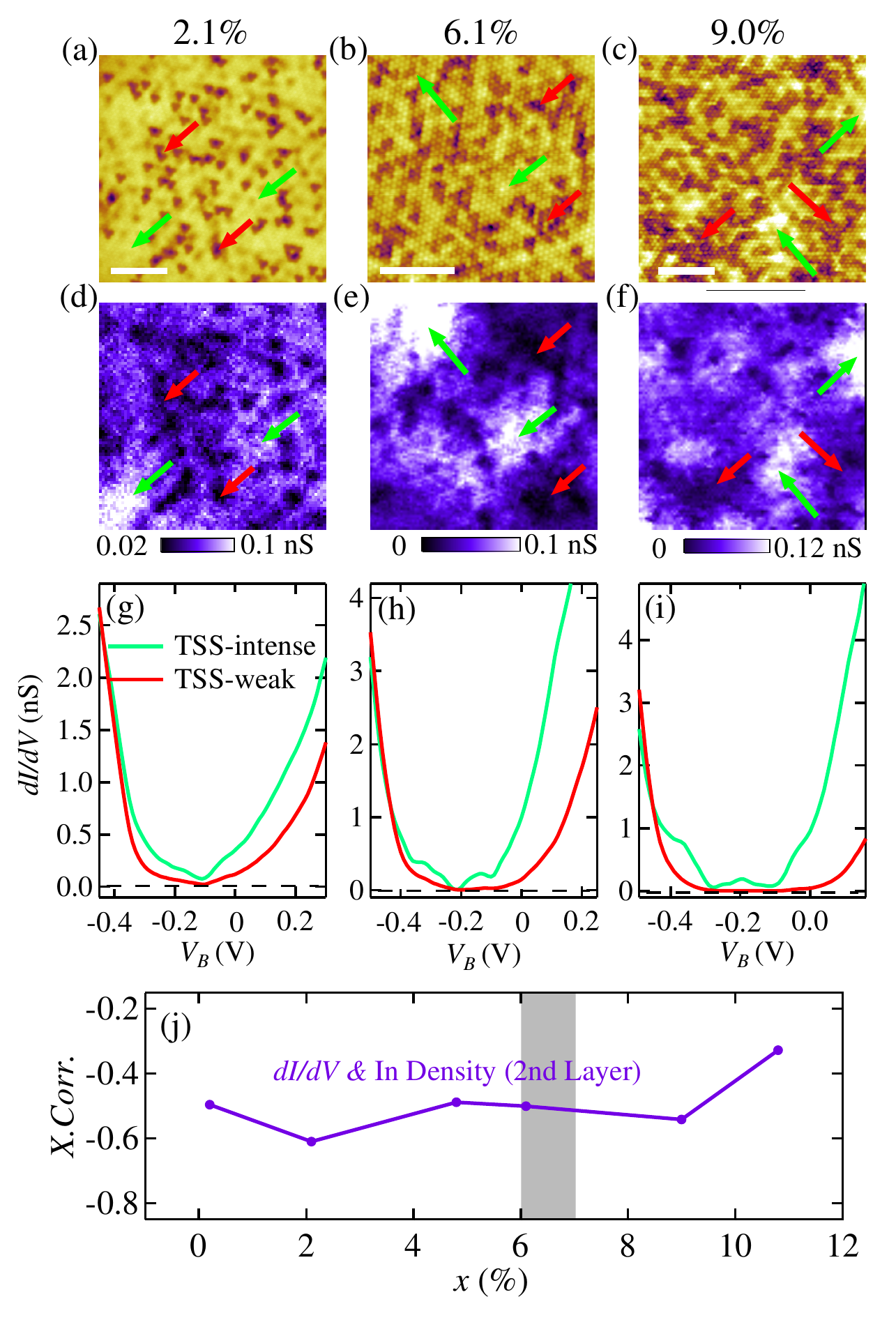}
\caption{{(a-c) Topographic images of $x= 2.1\%$, $6.1\%$ and $9.0\%$ (Tunneling condition: $-0.6$~V, 1~nA.). Scale bars are 5~nm. (d-f) $dI/dV$ maps in the same area of (a-c). When $x\leq x_C$, $E=E_D+0.05$~eV; for $x> x_C$, $E$ is inside the band gap, as indicated in Fig.\,\ref{fig1}(e). Red (green) arrows mark typical TSS-weak (intense) regions. (g-i) average $dI/dV$ spectra of selective regions: Red (green) is from TSS-weak (intense) area. (j) $x$ dependence of cross correlations between $dI/dV(\boldsymbol{r})$ and local density $n(\boldsymbol{r})$ of In in the second atomic layer.}}
\label{fig2}
\end{figure}

Fig.\,\ref{fig2} shows the spectroscopy mapping results on three representative samples across the TPT ($2.1\%$, $6.1\%$ and $9.0\%$). Fig.\,\ref{fig2}(a-c) show the topographic images where STS data were collected. In defects distribute randomly in the crystals. The $dI/dV$ maps of in-gap states (presumably the TSS) at the same locations are shown in Fig.\,\ref{fig2}(d-f). The $dI/dV$ maps of $x\leq 6.1\%$ are taken at $E=E_D+0.05$~eV to enhance the contrast. Note that the inhomogeneity of $dI/dV$ intensity is robust within the band gap. For $x> x_C$, the energy of the in-gap states (likely residue TSS) is marked by the arrow in Fig.\,\ref{fig1}(e). Clearly, the TSS is spatially inhomogeneous at nanometer scale. The average $dI/dV$ spectra of regions with weak (red arrows) and intense (green arrows) spectro-weight of TSS  are shown in Fig.\,\ref{fig2}(g-i). (See Supporting Information for details) In the TSS-intense area, the in-gap LDOS persists even above $x_C$ samples. In contrast, the in-gap LDOS of TSS-weak area decreases rapidly to zero as $x$ rises. The linear dispersion of TSS is visible only when $x<x_C$. More interestingly, the TSS anti-correlates with the local In density $n(\boldsymbol{r})$ (counting only those in second layer) as shown  by the negative cross correlation coefficients ($X.Corr.$) in Fig.\,\ref{fig2}(j). Note that it is practically impossible to count In defects at deeper layers. The observed anti-correlations indicate that  In defects are very effective ``suppressor'' of topological band inversion~\cite{Liu2013}. Thus, it is imperative to  reveal their impacts  individual In defects on local topological properties.

\begin{figure}[ht]
\centering
\includegraphics[width=0.5\columnwidth]{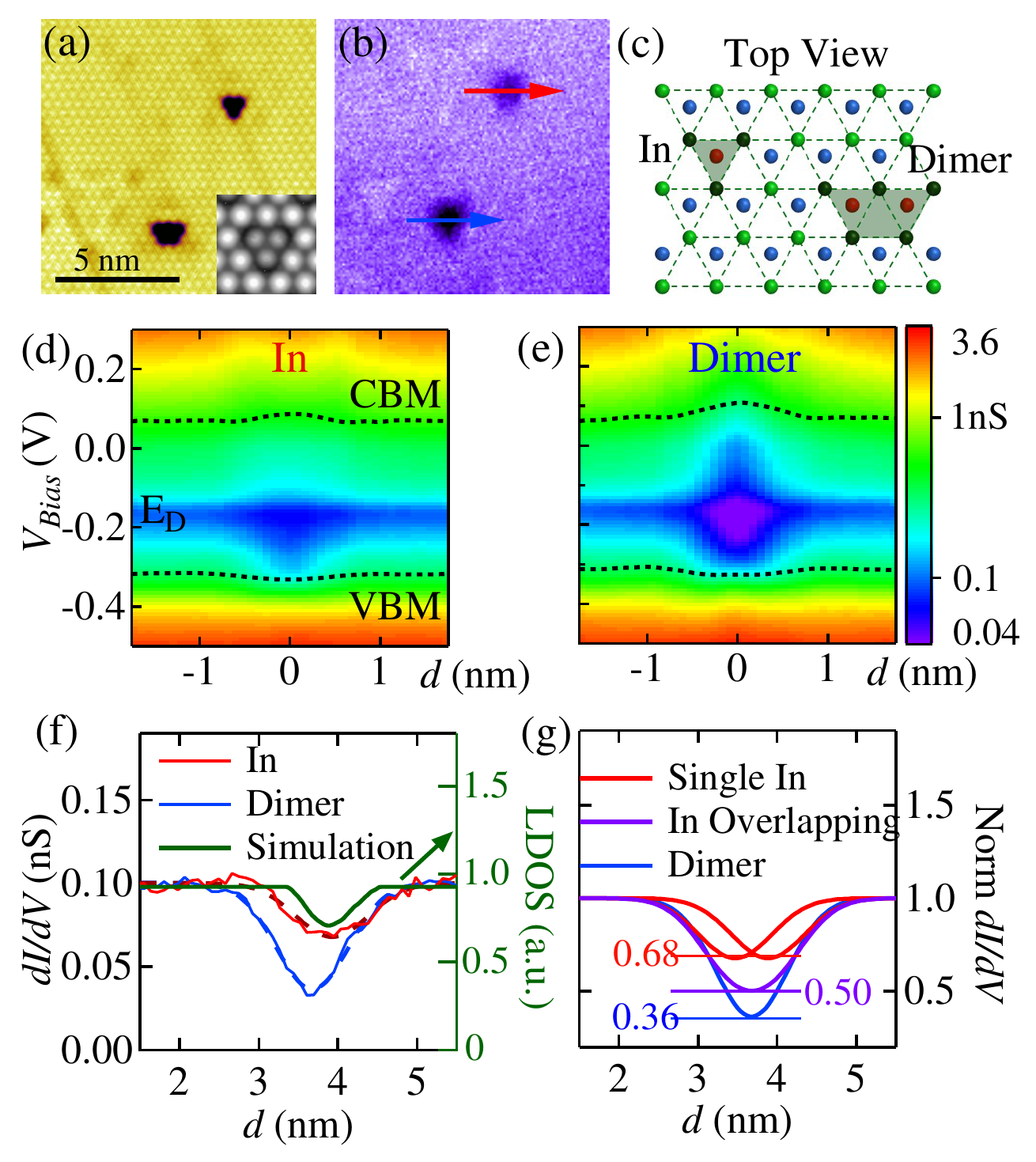}
\caption{(a) Topographic image of a single In and an In dimer on the second layer. Inset: simulated STM image of In defect by DFT. (b) STS map at $E_D$ in the same area as (a). (c) The top view of atomic structure of the top 2 atomic layers: Se (green), Bi (blue) and In (red). (d,e) $dI/dV$ intensity plots along the lines across single In (red) and In Dimer (blue). Tunneling conditions: $-0.6$\,V, 0.5\,nA. (f) Suppression of TSS at $E_D$ by In defects: blue, $dI/dV$ line profile across a In dimer; red, $dI/dV$ line profile across a single In defect; green: calculated $dI/dV$ of a single In defect. (g) Normalized Gaussian fitted of the $dI/dV$ spectra in (f): blue for In dimer; red for single In defects; purple for simulated overlapping of $dI/dV$ suppression by two single In defects.}
\label{fig3}
\end{figure}

Fig.\,\ref{fig3}(a) shows the typical topographic image of a single In defect and an In dimer with nearest neighboring In defects on the $0.2\%$ sample. Their configurations are illustrated in the schematic  in Fig.\,\ref{fig3}(c).  Fig.\,\ref{fig3}(b) shows a $dI/dV$ map measured at $E_D$ in the same area. Evidently the much lower spectro-weight on top of the In sites demonstrates that the TSS are strongly ``suppressed'' by In defects. This is further illustrated by the 2D spectral map (bias vs. displacement) of a $dI/dV$ line profile across the center of a single In defect (red arrow in Fig.\,\ref{fig3}(b)) shown in Fig.\,\ref{fig3}(d). In addition,  there is a slight increase of local band gap on the In defect, illustrated by the band edges (dash lines). The $dI/dV$ line profile at $E_D$ (red curves in Fig.\,\ref{fig3}(f)) shows a bell-shape suppression ($\sim$ 68\%) with the full width half maximum (FWHM) $\xi\approx1.17\pm0.03$\,nm, indicating the influence range of single In defects extends to approximately $3a$ ($a\approx0.41$\,nm is the in-plane lattice constant). Interestingly, the FWHM $\xi$ is comparable with the decay length of TSS, $\displaystyle\xi_S =\frac{\hbar v_F}{\Delta}\approx 0.94$\,nm, where $v_F\approx 5\times10^5$\,m/s is the Fermi velocity of Dirac surface states and $\Delta \approx 0.35$\,eV is the band gap~\cite{Xia2009}. The significant suppression of TSS  indicates that a single In defect can be approximated as a point NI embedded in the TI matrix.

More interestingly, the neighboring of In defects significantly enhances the  suppression of local topological band inversion. Fig.\,\ref{fig3}(e) displays the 2D spectral map of $dI/dV$ spectra taken along the line across an In dimer (blue arrow in Fig.\,\ref{fig3}(b)). In addition to the further enhancement of local band gap, the suppression of TSS on In dimer is much stronger than that on single In defect, and even stronger than a simple superposition of the suppression from two independent In defects. As shown in Fig.\,\ref{fig3}(g),  the simple overlapping of two single In defects  suppress the spectro-weight to 50\%, while the observed spectro-weight of In dimer is reduced to $\sim36\%$. Systematic studies of In dimers with different spacing indicates that such enhancement persists for In dimers with two In are separated by $3a$, consistent with the influence range $\xi$ of single In defects (see Supporting Information).  

DFT calculations were carried out to corroborate STM observation. The inset of Fig.\,\ref{fig3}(a) shows a simulated STM image of single In defect, in good agreement with STM results. Furthermore, the simulated LDOS of TSS near In defects shows a similar suppression (green curve in Fig.\,\ref{fig3}(f)), which is also in good agreement with experimental results. Because the TSS are protected by the band topology (the surface-bulk correspondence), they cannot be ``annihilated'' without changing the band topology\cite{Kane2010RMP}. Thus, the local suppression of TSS spectro-weight likely comes from an effective increase of the tunneling barrier width.  Because In defects convert nano-regions around them to NI, the interface between TI and NI (vacuum) is shifted slightly into the bulk, effectively increasing the tunneling barrier width and thus reducing the tunneling spectro-weight of TSS.

\begin{figure}[ht]
\centering
\includegraphics[width=0.5\columnwidth]{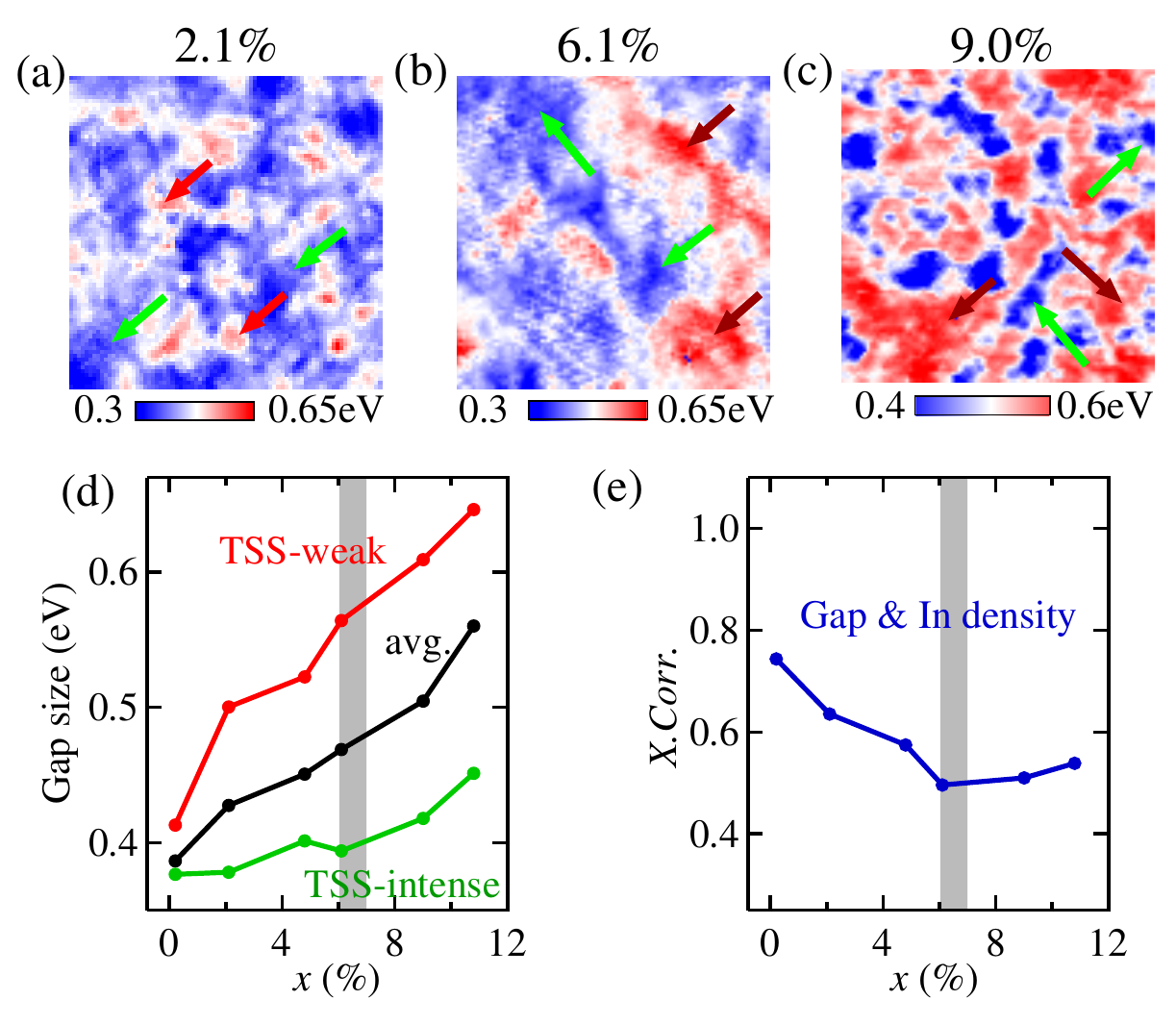}
\caption{(a-c) Local band gap maps $\Delta(\boldsymbol{r})$ of the same area as Fig.\,\ref{fig1}(a-c). (d)  Average band gaps of the whole (black), TSS-intense (green) and TSS-weak area (red) as functions of $x$. (e) $x$ dependence of the cross correlation coefficients between local In density $n(\boldsymbol{r})$ and band gap $\Delta(\boldsymbol{r})$.}
\label{fig4}
\end{figure} 

The observation of enhanced band gap $\Delta$ on In defects suggest a positive correlation between local In density and local band gap $\Delta(\boldsymbol{r})$ (extracted from spectroscopy maps using the method mentioned in Fig.\,\ref{fig1}).  Similar to the TSS, $\Delta(\boldsymbol{r})$ is also spatially inhomogeneous on nanometer scale, as shown in Fig.\,\ref{fig4}(a-c) (see complete data set in Support Information). Fig.\,\ref{fig4}(d) shows the $x$ dependence of average band gap $\left<\Delta\right>$ of selected regions. $\Delta(\boldsymbol{r})$ in the TSS-weak area (red arrows) is larger than that in the TSS-intense area (green arrows). $\left<\Delta\right>$ of TSS-weak regions increases rapidly while that of TSS-intense area changes little. Thus, the rise of $\left<\Delta\right>$ of the whole area (black curve) is mainly due to the increasing areal fraction of TSS-weak regions. Note that $\left<\Delta\right>$ of the whole area agrees well with that extracted from spatially averaged $dI/dV$ spectra in Fig.\,\ref{fig1}(d) (see Supporting Information), corroborating the validity of the band edge estimation method. Furthermore, $\Delta(\boldsymbol{r})$ is correlated with local In density $n(\boldsymbol{r})$, as shown by the positive cross correlation coefficients ($X.Corr.$) in Fig.\,\ref{fig4}(e).

\begin{figure}[ht]
\centering
\includegraphics[width=0.5\columnwidth]{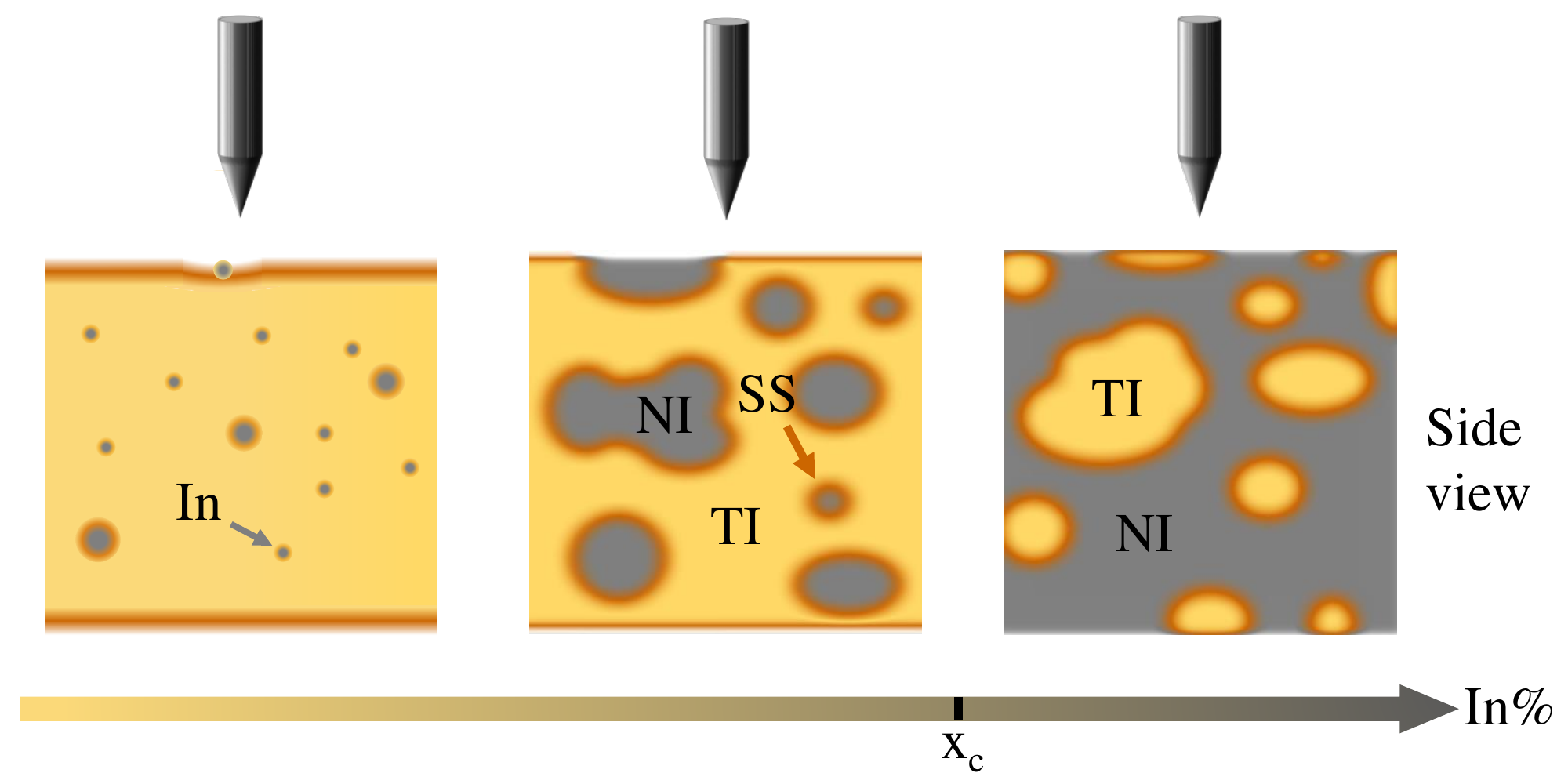}
\caption{Cartoon of the TPT due to nanoscale electronic inhomogeneity. The top and bottom are the boundaries of TI and vacuum. Left: In-dilute; Middle: underdoped, where topological regions dominate; Right: over-doped, where normal regions dominate.}
\label{fig5}
\end{figure}

The observed nanoscale electronic inhomogeneity of TSS and band gap in In doped Bi$_2$Se$_3$ may be understood by a  scenario illustrated in Fig.\,\ref{fig5} \color{black}In defects act as local topological-state ``suppressors'', converting nano-regions around them to NI. For In defects near the surface, the normal nano-regions (NNR) push the TSS slightly into the bulk, as illustrated in the left panel in Fig.\,\ref{fig5}, so that the tunneling spectral weight of TSS is suppressed in surface-sensitive measurements~\cite{Lou2015}.  As $x$ increases but $<x_c$, more regions with higher local In density form NNRs with larger band gaps in the matrix of topological regions which remain topological. For $x>x_c$, the topological regions form nano-size bubbles, denoted as topological nano-regions (TNRs) while the normal regions form the matrix. So the material is effectively a normal insulator (the right panel in Fig.\,\ref{fig5}). This phenomenological scenario qualitatively explains the nanoscale electronic inhomogeneity observed in our STM measurements. Further studies by other experimental techniques may help to better understand the unconventional TPT in  In doped Bi$_2$Se$_3$.

It is worth noting that this scenario does not contradict the band-closure scenario of TPT, which assumes spatially uniform SOC so the band gap must close at $x_c$. The assumption of spatially uniform SOC is invalid for (Bi$_{1-x}$In$_x$)$_2$Se$_3$. Prior DFT studies suggest that In 5$s$ orbitals are very effective on suprressing SOC, thus In defects can revert the band inversion locally~\cite{Liu2013}. Therefore,  In doping would inevitably introduce spatial SOC disorders, likely resulting in a mixture of TNRs and NNRs, a proliferation of TSS inside the bulk crystal, and a gradual increase of average band gap across the TPT~\cite{Lou2015}. The proliferation of TSS inside the bulk crystal is also consistent with the large enhancement of optical absorption around $x_c$~\cite{Wu2013a}. Note that the inhomogeneous TPT scenario present here is different from the theoretical proposal of bypassing band gap closure via symmetry broken states\cite{Ezawa2013}.

To conclude, our results provide compelling microscopic evidence of an inhomogeneous TPT in (Bi$_{1-x}$In$_x$)$_2$Se$_3$, which is driven by nanoscale mixture of NNRs and TNRs. As shown by our systematic STM and STS studies, the inhomogeneity of both TSS and local band gap originates from very effective suppression of local SOC (and band topology) by In defects, resulting in local NNRs. Our results demonstrate that strong disorders can play a significant role in  TPT, which is difficult to capture in spatially average measurements. The direct observation of nanoscale inhomogeneous TPT will motivate further studies on the impact of disorders in the topological materials and the associated quantum phase transitions.

\begin{acknowledgement}

We are grateful to J. Liu, D. Vanvderbilt, P. Armitage and S. Oh for helpful discussions. The STM work was supported by NSF under Grand No. DMR-1506618. The synthesis work was supported by the NSF under Grant No. DMR-1629059. The theoretical work is supported by the National Natural Science Foundation of China (grant No. 11774084). Z. Zhong acknowledges financial support by CAS Pioneer Hundred Talents Program.

\end{acknowledgement}

\begin{suppinfo}
High-resolution Topographic Images of (Bi$_{1-x}$In$_x$)$_2$Se$_3$.
\newline
Simulation of the Evolution of DOS with Band Closure.
\newline
Estimation of Band Edges via Linear Curve fitting on Average $dI/dV$ Spectrum.
\newline
Estimated Band Gap via $d^{2}I/dV^{2}$ Spectra.
\newline
Definition of TSS-weak and TSS-intense Area and the Corresponding $dI/dV$ Spectra.
\newline
Anticorrelation between the Spectral Weight of the Topological Surface States and Local In Density $n(\boldsymbol{r})$.
\newline
Suppression of Surface States by Single In Defect and In Dimers.
\newline
Comparison of Estimated Band Edges from Averaged $dI/dV$ Spectra and Averaged Band Edges from $dI/dV$ Maps.
\newline
Band Structure Evolution across the Topological Phase Transition.
\newline
\end{suppinfo}


\end{document}